\documentclass[twocolumn,prb,showpacs,multicol,amsmath,amssymb]{revtex4-2}
\usepackage{amssymb}
\usepackage{amsmath}
\usepackage{graphicx}
\usepackage{epstopdf}
\usepackage{bm}
\usepackage{gensymb}

\newcommand{\be}{\begin{equation}}
	\newcommand{\ee}{\end{equation}}

\newcommand{\bea}{\begin{eqnarray}}
	\newcommand{\eea}{\end{eqnarray}}

\newcommand{\bd}{\begin{displaymath}}
	\newcommand{\ed}{\end{displaymath}}
\newcommand{\ba}{\begin{array}}
	\newcommand{\ea}{\end{array}}
\newcommand{\bi}{\begin{itemize}}
	\newcommand{\ei}{\end{itemize}}
\newcommand{\bc}{\begin{center}}
	\newcommand{\ec}{\end{center}}
\newcommand{\bfl}{\begin{flushleft}}
	\newcommand{\efl}{\end{flushleft}}
\newcommand{\bfr}{\begin{flushright}}
	\newcommand{\efr}{\end{flushright}}

  \def\bd{{\bf d}}

\def\6{\partial}

\def\={\!\!\!&=&\!\!\!}
\def\+{\!\!\!&&\!\!\!+~}
\def\-{\!\!\!&&\!\!\!-~}

\newcommand\redout{\bgroup\markoverwith{\textcolor{red}{\rule[.5ex]{2pt}{0.4pt}}}\ULon}

\graphicspath{{.}{./plot/}}

\begin{document}
	
	\title{Mechanism of Incommensurate Magnetic Order in $\mathrm{BaCo_2(AsO_4)_2}$: Interplay of Frustrated Further-Neighbor Exchanges and Bond-Directional Anisotropy}
	
\date{\today}

\author{Mohammad-Hossein Zare}\email{zare@qut.ac.ir}
\affiliation{Department of Physics, Qom University of Technology, Qom 37181-46645, Iran}
\author{Mehdi Biderang}
\affiliation{Department of Physics, University of Toronto, 60 St. George Street, Toronto, Ontario M5S 1A7, Canada}
\author{Hamid Mosadeq}\email{mosadegh@sku.ac.ir}
\affiliation{Department of Physics, Faculty of Science, Shahrekord University, Shahrekord 88186-34141, Iran }

\begin{abstract}
The microscopic mechanism governing the zero-field incommensurate magnetic order in the honeycomb cobaltate $\mathrm{BaCo}_2(\mathrm{AsO}_4)_2$ remains a significant unresolved problem, particularly concerning the relative contributions of bond-directional Kitaev-type interactions and exchange frustration. This study investigates an extended $J_1\text{--}K\text{--}\Gamma\text{--}\Gamma'\text{--}J_2\text{--}J_3$ model, which incorporates XXZ-type exchange anisotropies on the honeycomb lattice. Nearest-neighbor parameters are constrained by ab initio electronic structure calculations. By integrating the analytical Luttinger-Tisza approach with exact diagonalization calculations under twisted boundary conditions, we delineate the classical and quantum phase diagrams across the $(J_2, J_3)$ parameter space. We demonstrate that the stabilization of the experimentally observed incommensurate spiral phase, propagating along the $\Gamma\to M$ direction, does not necessitate an anomalously dominant Kitaev coupling. Instead, this phase arises naturally from the synergistic interplay between exchange frustration, driven by further-neighbor couplings ($J_2, J_3$), and intermediate off-diagonal bond anisotropies ($\Gamma, \Gamma'$). In the quantum regime, we elucidate the competition between this incommensurate manifold and an out-of-plane ferromagnetic ($\mathrm{FM}_z$) phase, which is selectively stabilized via a quantum order-by-disorder mechanism. Our findings reconcile conflicting interpretations of the magnetic interactions in $\mathrm{BaCo}_2(\mathrm{AsO}_4)_2$ and establish the microscopic origin and stability range of its incommensurate ground state.

\end{abstract}

\maketitle

\section{introduction}
Two-dimensional (2D) honeycomb lattice magnetic materials are of intense interest due to the Kitaev spin model, which spin-$S=1/2$ moments on the honeycomb lattice are coupled with bond-dependent frustrating interactions, had the potential in realizing a quantum spin liquid (QSL)~\cite{Kitaev:2006}.
This exotic state of matter is formed by creating a collective quantum state with long-ranged  entangled spins, topological ordering, fractionalized spin excitations, and the absence of spontaneous symmetry breaking down to the lowest temperatures~\cite{Balents:2010,Balents:2016,Broholm:2020}.
The Kitaev spin model is exactly solvable and gives a QSL ground state host topologically protected fractionalized excitations~\cite{Kitaev:2006}.
The exotic excitations of the Kitaev QSL are known as Majorana fermions and $\mathbb{Z}_2$ gauge fluxes~\cite{Zhou:2017,Hermanns:2018,Knolle:2019,Motome:2020}, may serve as the basic primitive of topological quantum computing~\cite{KITAEV:20032,Nayak:2008,Ahlbrecht:2009}.

Material realizations of this model have been proposed in 4d- or 5d-heavy transition metal compounds with edge-sharing octahedra in a honeycomb lattice, where strong spin-orbit coupling competes with electron correlation~\cite{Jackeli:2009}.
In all real materials are emerged as candidates for the realization of the Kitaev QSL state~\cite{Rau:2014}, Na$_2$IrO$_3$~\cite{Singh:2010,Singh:2012}, $\alpha$-RuCl$_3$~\cite{Plumb:2014,Banerjee:2016}, and $\alpha$-Li$_2$IrO$_3$~\cite{Kobayashi:2033}, all magnetically order by non-Kitaev nearest neighbor interactions including Heisenberg interaction, $J$, and bond-dependent off-diagonal couplings $(\Gamma,~\Gamma')$ are unavoidable~\cite{Rau:2016,Takagi:2019}.
Although $4d$ honeycomb Mott insulator $\alpha$-RuCl$_3$ establishes a long-range antiferromagnetic zigzag (ZZ) order at low temperatures, it  is predicted to be proximate to the pure Kitaev points.
Theoretical and experimental efforts have indicated that the magnetic order is fragile and can be suppressed by applying an external magnetic field~\cite{Sears:2015,Johnson:2015,Kubota:2015,Janssen:2016,Yadav:2016,Leahy:2017,Baek:2017,Kasahara:2018,Banerjee:2018}, and an intermediate QSL state can be realized between the low-field and high-field phases for a finite range of magnetic field.
Recent experimental observations provide strong evidence for the Majorana fermion as origin of the half-integer quantization plateau of thermal Hall conductivity in the $4d$-based Kitev magnet $\alpha$-RuCl$_3$~\cite{Banerjee:2016,Banerjee:2017,Banerjee:2018}. 

More recently, the focus of intensive research is devoted to build the Kitaev-dominant systems in materials with less extended $3d$ orbitals~\cite{Liu:2018,Sano:2018}.
Several Co-based transition metal compounds with a honeycomb arrangement of 
$3d^7$ Co$^{+2}$ ions such as BaCo$_2$(AsO$_4$)$_2$ (BCAO)~\cite{REGNAULT:1977,REGNAULT:2018,REGNAULT:1979},  Na$_2$Co$_2$TeO$_6$~\cite{VICIU:2007,Lefran:2016}, Na$_3$Co$_2$SbO$_6$~\cite{VICIU:2007,Yan:2019}, and BaCo$_2$(PO$_4$)$_2$~\cite{Nair:2018} have been proposed as promising candidate materials to host the exotic Kitaev QSL state.
These Kitaev model candidate materials exhibit conventional long-range or short-range magnetic ordering at low temperature due to additional non-Kitaev interactions.
Most recent theoretical and experimental effort have argued that the magnetic properties and excitations of these Co$^{+2}$ materials can be described in the context of the anisotropic XXZ Heisenberg model~\cite{Das:2021,Halloran:2023} and the Kitaev-type $JK\Gamma\Gamma'$ model.

The zero-field ordered state of honeycomb-layered cobaltates BCAO is an incommensurate antiferromagnetic (AFM) order with an ordering wavevector along the $\Gamma-{\rm M}$ high-symmetry line~\cite{REGNAULT:1979,Halloran:2023}.
However, recent experimental observations report that it hosts double-zigzag (dZZ) spin-chains form a $\uparrow\uparrow\downarrow\downarrow$ pattern with small out-of-plane canting angle~\cite{REGNAULT:2018,tu:2023}.
It is worth mentioning that the dZZ phase is close in energy to the incommensurate spiral structure.
The near degeneracy of these two frustrated incommensurate magnetic phases may be lifted by delicate effects such as quantum fluctuation or lattice distortions that could either favor the dZZ phase or the incommensurate order.
Here, the central question we address is how to theoretically realize the incommensurate spiral order along the $\Gamma-{\rm M}$ high-symmetry line (labelled as $I_{\Gamma-{\rm M}}$).

The detailed microscopic origin of the incommensurate spiral phase remains
unknown due to the absence of an effective spin Hamiltonian for BCAO.
In this study, in order to address this question, we utilize analytical and numerical methods on the extended Heisenberg-Kitaev-Gamma-Gamma$'$ ($JK\Gamma\Gamma'$) model with the significant nearest-neighbor Heisenberg interaction of the $3d^7$ honeycomb cobaltate BCAO~\cite{Liu:2023}
to investigate the emergence of the incommensurate magnetic order with the ordering wavevector between the $\Gamma$ and M points.
Experimental and theoretical works reported that trigonal lattice distortion plays a key role in the physics of BCAO owing to the isotropic Heisenberg interaction $J$ becomes the $XXZ$ type~\cite{Halloran:2023,Das:2021,Winter:2022,Maksimov:2022,Liu:2023}.
Our calculations indicate that both the second- and third-NN Heisenberg exchange interactions in the presence of anisotropic contribute to the formation of the incommensurate magnetic structure of BCAO.  

Recent experimental and theoretical advancements have necessitated a fundamental re-evaluation of the microscopic physics governing the honeycomb magnet BCAO. Historically, BCAO has been regarded as a model realization of the two-dimensional $XY$ system; however, recent inelastic neutron scattering data obtained in the field-polarized phase have revealed a significant energy splitting ($\Delta E_{\rm M}$) at the M point of the Brillouin zone when comparing different in-plane field orientations ($B \parallel b$ and $B \parallel a^*$) \cite{maksimov2025strong}. This observation provides unequivocal evidence for dominant Kitaev interactions, establishing BCAO as a Kitaev champion among $3d^7$ cobaltates, with bond-dependent exchanges nearly twice the magnitude of the Heisenberg term \cite{maksimov2025strong, Halloran:2023}. Moreover, the nature of the zero-field ground state has been clarified as a non-classical dZZ order, characterized by a distinct out-of-plane spin tilt, which defies classical spiral predictions \cite{maksimov2025strong, devillez2025bond}. The stabilization of this collinear phase against competing spiral states is attributed to a robust quantum order-by-disorder mechanism, wherein quantum fluctuations lift classical degeneracies to favor specific ordered arrangements \cite{lee2025quantum, chernyshev2025quantum}. Notably, ultra-low temperature thermal transport measurements have provided the first direct evidence of a field-induced gapless QSL phase in BCAO \cite{tu:2023}. The detection of a finite residual linear term in thermal conductivity ($\kappa_0/T$) at intermediate fields ($\sim 0.5$meV) of the $\text{CoO}6$ octahedra, which causes the $J_{\text{eff}}=1/2$ pseudospin states to deviate significantly from the ideal limit ($J_{\text{eff}} \approx 1/2$) and profoundly renormalizes the anisotropic exchange manifold \cite{ferreira2025trigonal, samanta2024spin}.

The remainder of this paper is organized as follows. Section II introduces the extended $J_1$-$K$-$\Gamma$-$\Gamma'$-$J_2$-$J_3$ spin Hamiltonian with XXZ-type exchange anisotropies on the honeycomb lattice and presents the classical phase diagrams obtained via the analytical Luttinger-Tisza (LT) approach. Section III describes the exact diagonalization (ED) framework formulated on an $N=18$ site cluster subject to continuous twisted boundary conditions TBCs. Section IV is devoted to the quantum phase diagram and ground-state magnetic properties, elucidating the stabilization of the incommensurate spiral state, the quantum order-by-disorder mechanism, and the evolution of the characteristic wavevector. Finally, Section V summarizes our primary conclusions and discusses their broader implications for candidate honeycomb materials.

\section{MODEL AND CLASSICAL PHASE DIAGRAM}

To model the highly competitive exchange interactions in the cobalt-based honeycomb magnet BCAO, we employ an extended Heisenberg-Kitaev-$\Gamma$-$\Gamma'$  ($JK\Gamma\Gamma'$) Hamiltonian on a honeycomb lattice. The local environment of the $\text{Co}^{2+}$ ions induces strong spin-orbit entanglement, which gives rise to bond-directional anisotropic exchange couplings alongside conventional isotropic interactions. The total classical Hamiltonian is defined as:

\begin{align}
	\mathcal{H}& = \sum_{\langle i,j \rangle_\gamma} \Big[ J_1 (S_{i}^{x}S_{j}^{x}+S_{i}^{y}S_{j}^{y}+\Delta_1 S_{i}^{z}S_{j}^{z}) \nonumber \\ & + K S_i^\gamma S_j^\gamma 
	+ \Gamma \left( S_i^\alpha S_j^\beta + S_i^\beta S_j^\alpha \right) 	\nonumber \\
	&  + \Gamma' \left( S_i^\gamma S_j^\alpha + S_i^\gamma S_j^\beta + S_i^\alpha S_j^\gamma + S_i^\beta S_j^\gamma \right) \Big] 	\nonumber \\
	&  
+ \sum_{\langle\langle i,j \rangle\rangle} J_2 S_i \cdot S_j+  \sum_{\langle\langle\langle i,j \rangle\rangle\rangle} J_3 (S_{i}^{x}S_{j}^{x}+S_{i}^{y}S_{j}^{y}+\Delta_3 S_{i}^{z}S_{j}^{z}),
	\label{eq:Hamiltonian}
\end{align}
where $\langle i,j \rangle_\gamma$ denotes a pair of nearest neighbors linked by a $\gamma$-bond ($\gamma \in \{x,y,z\}$), and $\{\alpha,\beta\}$ represents the two Cartesian spin components orthogonal to $\gamma$. The parameters $J_1$, $K$, $\Gamma$, and $\Gamma'$ correspond to the Heisenberg, Kitaev, symmetric off-diagonal, and pseudo-dipolar off-diagonal exchange energies, respectively. Furthermore, $J_2$ and $J_3$ denote the second- and third-nearest-neighbor Heisenberg interactions. Notably, both the first- and third-nearest-neighbor exchange couplings incorporate spatial XXZ anisotropy, parameterized by $0 \le \Delta_n \le 1$ (for $n=1,3$). 

Utilizing the tight-binding parameters extracted for BCAO via ab initio electronic structure calculations~\cite{Liu:2023}, the realistic set of exchange interactions (in units of meV) under trigonal distortion is parameterized as:
\begin{equation}
	(J_1, K, \Gamma, \Gamma') = (-8.14, -0.47, 2.63, 2.85),
	\label{eq:BCAO_parameters}
\end{equation}
while treating the second- and third-nearest-neighbor couplings $J_2$, $J_3$, and the XXZ anisotropy factor $\Delta_n$ as open parameters whose individual and competing roles in driving or protecting magnetic stability are to be critically evaluated. Recent experimental investigations on cobalt-based $d^7$ honeycomb magnets, most notably BCAO, have unveiled a highly anisotropic magnetic response~\cite{REGNAULT:1979,Halloran:2023}. This distinctive feature has renewed intensive theoretical interest in highly anisotropic XXZ variants of honeycomb lattice. This framework allows us to systematically decipher the microscopic mechanisms driving the incommensurate magnetic ordering of BCAO and discriminate which of these key parameters- $J_2$, $J_3$, or $\Delta_n$- plays the predominant role in stabilizing the underlying incommensurate phase against competing states.

\begin{figure}[t]
	\includegraphics[width=3.60in]{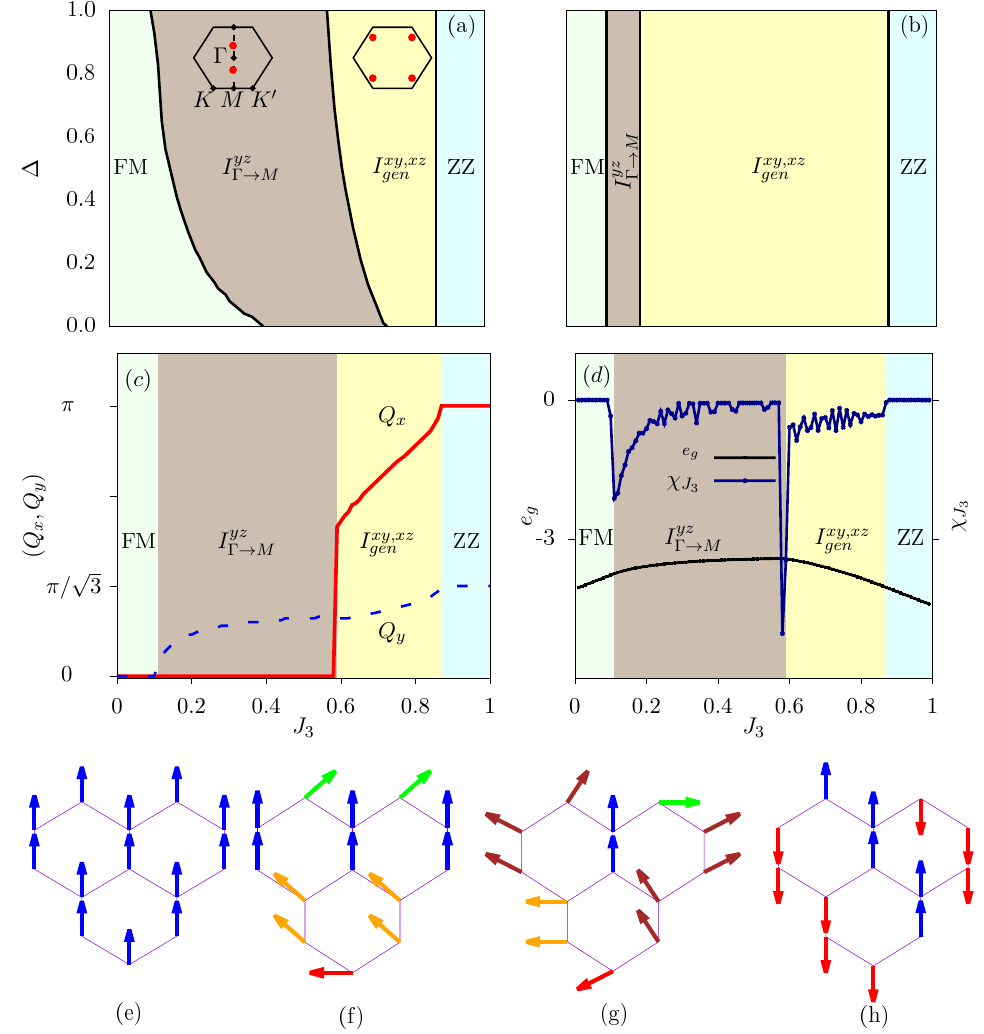}
	\caption{ 	
		Classical phase diagrams and magnetic instabilities of the extended honeycomb $JK\Gamma\Gamma'$ model at $J_2 = 0$. 
		(a) Phase diagram in the $J_3$--$\Delta$ plane for the partial-XXZ $J_1K\Gamma\Gamma'J_3^\Delta$ model, with isotropic couplings fixed to the ab initio values of BCAO [Eq.~\eqref{eq:BCAO_parameters}]. Inset: The FBZ with high-symmetry points and propagation paths for $\mathbf{Q}$. 
		(b) Symmetrical phase diagrams for the $J_1^\Delta K\Gamma\Gamma'J_3$ and full-XXZ $J_1^\Delta K\Gamma\Gamma'J_3^\Delta$ models. 
		(c) Evolution of the ordering vector $\mathbf{Q}$ versus $J_3$ for the $J_1K\Gamma\Gamma'J_3^{\Delta}$ model at anisotropy $\Delta = 0.1$. 
		(d) Ground-state energy per site $e_g$ and classical susceptibility $\chi_{J_3} = -\partial^2 e_g / \partial J_3^2$ versus $J_3$ at $\Delta = 0.1$, highlighting sharp first-order anomalies and broad second-order transitions. 
		(e)--(h) Real-space spin configurations of the ground states: (e) FM order, (f) the coplanar incommensurate $I_{\Gamma\rightarrow {\rm M}}^{yz}$ state with $\mathbf{Q}=(0,\pi/(2\sqrt{3}))$, (g) the doubly degenerate $I_{\rm gen}^{xy(xz)}$ spiral state $\mathbf{Q}=(\pi/3,2\pi/(3\sqrt{3}))$, and (h) the ZZ order.}
	\label{fig:j2d}
\end{figure}

\subsection{Luttinger-Tisza Optimization and Fourier Representation}
\label{subsec:luttinger_tisza}

To establish a baseline for magnetic instabilities, the classical phase diagram of Eq. \eqref{eq:Hamiltonian} is analyzed using the LT method \cite{Luttinger:1946,LITVIN:1974,Sklan:2013,Zare:2013,Kimchi:2014,Chaloupka:2015,Nishimoto:2016,Zare:2021}. This approach replaces the rigid local spin constraint, $|S_i|^2 = S^2$ at each lattice site $i$, with a relaxed, global constraint summed over the entire system: $\sum_i |S_i|^2 = N S^2$. By transforming the classical spin degrees of freedom into momentum space via a multi-sublattice Fourier definition, the Hamiltonian is expressed in a quadratic form:

\begin{equation}
	\mathcal{H} = \sum_{\mathbf{q}} \Psi^\dagger(\mathbf{q}) \, \mathbb{M}(\mathbf{q}) \, \Psi(\mathbf{q}),
	\label{eq:LT_fourier}
\end{equation}
where $\Psi(\mathbf{q}) = \left( \mathbf{S}_A(\mathbf{q}), \mathbf{S}_B(\mathbf{q}) \right)^T$ is a six-component column vector representing the Fourier components of the spins on the two interpenetrating triangular sublattices ($A$ and $B$) of the honeycomb network, and $\mathbb{M}(\mathbf{q})$ is a $6 \times 6$ Hermitian interaction matrix that incorporates the directional bond couplings.

Under the weak constraint, the classical ground-state energy corresponds to the lowest eigenvalue of $\mathbb{M}(\mathbf{q})$, denoted as $\lambda_{\text{min}}(\mathbf{q})$, optimized across the entire first Brillouin zone. The wavevector $\mathbf{Q}$ that minimizes this energy profile ($\nabla_{\mathbf{q}} \lambda_{\text{min}}(\mathbf{q})|_{\mathbf{q}=\mathbf{Q}} = 0$) determines the spatial modulation of the ordering pattern. In regions where the corresponding eigenvector identically satisfies the local spin constraint across all sites, the LT solution provides the exact classical ground state.

\subsection{Classical Phase Diagram and Incommensurate Instabilities}

\label{subsec:classical_results}

\begin{figure}[t]
	\includegraphics[width=3.60in]{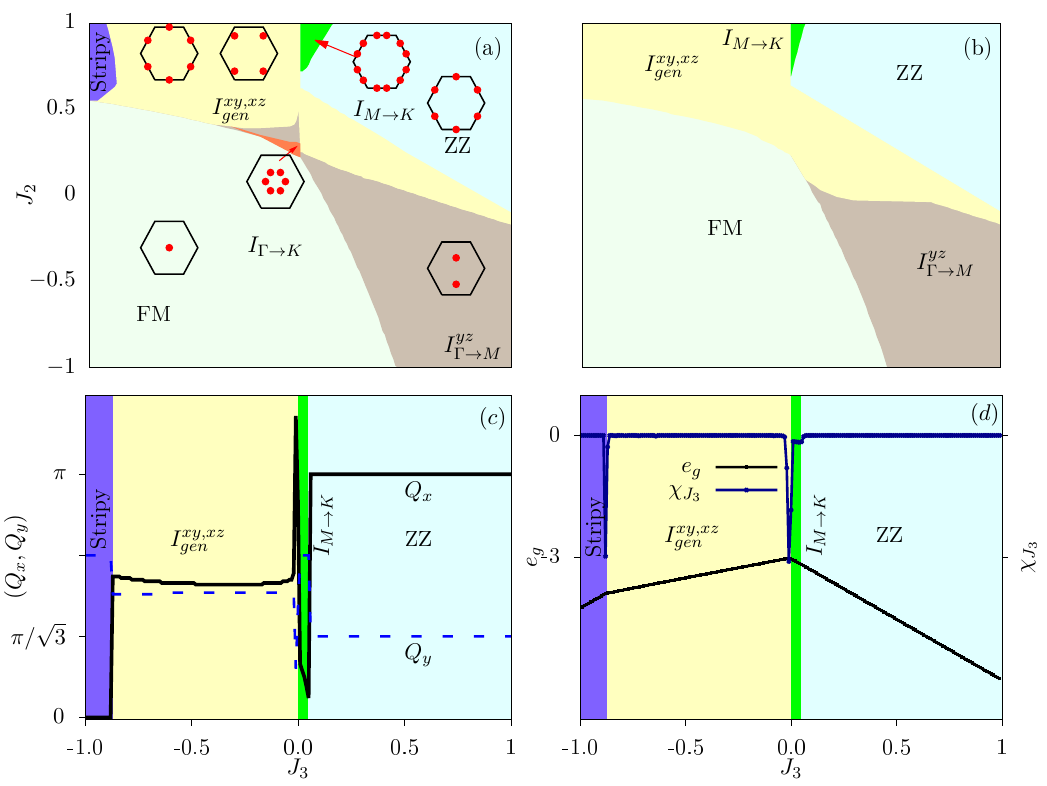}
	\caption{
		(Color online) 
		Classical magnetic phase diagrams and the evolution of modulated instabilities in the $J_3$--$J_2$ plane for the full-XXZ $J_1^\Delta K\Gamma\Gamma'J_3^\Delta$ model, derived from an ab initio baseline of BCAO, are presented. The phase landscapes are resolved at fixed XXZ anisotropy limits: (a) the isotropic Heisenberg limit ($\Delta = 1$) and (b) the pure planar limit ($\Delta = 0$). Panels (c) and (d) depict the evolution of the magnetic ordering wavevector $\mathbf{Q}$ and the classical ground-state energy per site $e_g$, alongside its susceptibility $\chi_{J_3} = -\partial^2 e_g / \partial J_3^2$ as a function of $J_3$, respectively. These are evaluated at fixed $J_2=0.4$ and $\Delta=0.8$ using the Luttinger-Tisza approach.
	}
	\label{fig:j3j2}
\end{figure}

The classical phase boundaries, determined via global minimization of the Fourier-transformed interaction matrix, reveal a rich landscape that distinctly differentiates between commensurate locking and continuously modulated, incommensurate textures across varying anisotropy regimes (as depicted in Figs.~\ref{fig:j2d} and \ref{fig:j3j2}). Notably, the classical ground states do not consistently revert to standard commensurate orders throughout the entire parameter space. Instead, even in proximity to the symmetric limits, a distinct portion of the phase diagram remains occupied by incommensurate spiral phases, specifically $I_{\Gamma\rightarrow {\rm M}}^{yz}$ and $I_{\rm gen}^{xy(xz)}$. This persistent unpinning of the ordering wavevector $\mathbf{Q}$ from the high-symmetry points of the first Brillouin zone (FBZ) is attributed to robust bond-dependent off-diagonal $\Gamma$ and $\Gamma'$ interactions.

The structural transition between distinct magnetic regimes is governed by the competition among these differing energy minima. The evolving landscape of these minimum-energy solutions and the resultant phase boundaries can be systematically monitored by examining the second derivative of the ground-state energy density ($e_g$) with respect to the coupling parameter $J_3$, which defines the classical magnetic susceptibility:
\begin{equation}
	\chi_{J_3} = -\frac{\partial^2 e_g}{\partial J_3^2}.
	\label{eq:susceptibility}
\end{equation}

To theoretically assess the definitive roles of the second- and third-nearest-neighbor couplings ($J_2, J_3$) in conjunction with the XXZ anisotropy ($\Delta_n$) in stabilizing the magnetic configurations of BCAO, we establish our baseline calculations using the realistic ab initio exchange parameters specified in Eq.~\eqref{eq:BCAO_parameters}. As a primary step toward understanding the stabilization and persistence of the incommensurate spiral configurations alongside standard conventional phases, we analyze the classical ground states in the $J_3$--$\Delta$ plane for a vanishing second-nearest-neighbor interaction ($J_2 = 0$), as depicted in Fig.~\ref{fig:j2d}. The critical importance of the third-nearest-neighbor exchange coupling in stabilizing highly competing magnetic structures in cobalt-based honeycomb magnets is well-established \cite{Maksimov:2022,Winter:2022,Halloran:2023,Nair:2018}.

Herein, we systematically investigate three distinct anisotropic variants of the underlying Hamiltonian:

\begin{enumerate}
	\item The partial-XXZ version where the $J_1$-term remains in the isotropic Heisenberg limit ($\Delta_1 = 1$) while the $J_3$-term incorporates variable anisotropy, designated as the $J_1K\Gamma\Gamma'J_3^\Delta$ model (as depicted in Fig.~\ref{fig:j2d}(a)).
	\item The alternative partial-XXZ variant where the $J_3$-term is isotropic ($\Delta_3 = 1$) and the $J_1$-term is anisotropic, denoted as the $J_1^\Delta K\Gamma\Gamma'J_3$ model (as depicted in Fig.~\ref{fig:j2d}(b)).
	\item The full-XXZ version characterized by uniform anisotropy across both configurations ($\Delta_1 = \Delta_3 = \Delta$), referred to as the $J_1^\Delta K\Gamma\Gamma'J_3^\Delta$ model (as depicted in Fig.~\ref{fig:j2d}(b)).
\end{enumerate}

The resultant classical magnetic phase diagrams in the $J_3$--$\Delta$ plane are evaluated for the fixed configuration $(J_1, K, \Gamma, \Gamma', J_2) = (-8.14, -0.47, 2.63, 2.85, 0)$, based on the ab initio parameters of BCAO. Our calculations reveal four distinct ordered regimes: two governed by commensurate wavevectors (the FM and ZZ phases) and two hosting incommensurate spin-spiral structures, which we label as $I_{\Gamma\rightarrow {\rm M}}^{yz}$ and $I_{\rm gen}^{xy(xz)}$, as shown in Figs.~\ref{fig:j2d}(e)--(h).

At the classical level, these incommensurate spiral orders, wherein the spins are strictly confined to the coplanar $yz$- and doubly degenerate $xy(xz)$-planes, respectively, are situated within a distinct regime separating the low-$J_3$ FM order and the high-$J_3$ ZZ phase. Since all relevant classical modes are coplanar, these phases exhibit remarkably stable behavior against variations in the spin anisotropy parameter $\Delta$. Our classical calculations for the $J_1K\Gamma\Gamma'J_3^\Delta$ model in Fig.~\ref{fig:j2d}(a) indicate that the incommensurate $\Gamma \to M$ ordering is robustly restricted to intermediate values of $J_3$. This underscores that while the $z$-axis component of the third-nearest-neighbor exchange interaction does not drastically expand the stabilization regime of the $I_{\Gamma\rightarrow {\rm M}}^{yz}$ phase, this modulated texture remains a persistent feature of the phase diagram.

The evolution of the ordering wavevector as a function of $J_3$ at a representative anisotropy of $\Delta = 0.1$ is depicted in Fig.~\ref{fig:j2d}(c). Energy minimization in reciprocal space reveals that the energy minimum remains locked at the $\Gamma$ point (the zone center, $\mathbf{Q}=0$) for weak third-nearest-neighbor antiferromagnetic couplings ($J_3 < 0.12$). For moderately strong interactions ($0.12 < J_3 < 0.6$), the minimum-energy solution continuously shifts along the $\Gamma-{\rm M}$ high-symmetry axis of the FBZ, signaling the emergence of the incommensurate $I_{\Gamma\rightarrow {\rm M}}^{yz}$ spiral phase. Within the interval $0.6 < J_3 < 0.85$, the system undergoes a spin-reorientation transition into the doubly degenerate $I_{\rm gen}^{xy(xz)}$ state near the boundary $J_3 \approx 0.6$. In this regime, the ordering wavevector $\mathbf{Q}$ unpins from the high-symmetry $\Gamma-{\rm M}$ line, migrating generically into the interior of the FBZ.  Hereafter, this is referred to as the generic incommensurate spiral phase ($I_{\text{gen}}$). It is noteworthy that the intrinsic sixfold degeneracy of the isolated planar spiral state is explicitly broken by the bond-dependent off-diagonal exchange interactions ($\Gamma, \Gamma'$) inherent in the model. Finally, for strong couplings ($J_3 > 0.85$), the system locks into the long-range ZZ order localized at the FBZ boundary ($\mathbf{Q}={\rm M}$).

Fig.~\ref{fig:j2d}(d) presents the classical ground-state energy per site ($e_g$) and its corresponding susceptibility, $\chi_{J_3}$. A precise correspondence is observed between the phase boundaries delineated in the classical phase diagrams and the loci of anomalous behavior in $\chi_{J_3}$. The broad, symmetric peaks in $\chi_{J_3}$ proximal to the outer phase boundaries are indicative of continuous, second-order phase transitions from the FM state to the $I_{\Gamma\rightarrow {\rm M}}^{yz}$ phase, and from the $I_{\rm gen}^{xy(xz)}$ state to the ZZ phase. Notably, the susceptibility anomaly associated with the transition delineating the $I_{\Gamma\rightarrow {\rm M}}^{yz}$ and $I_{\rm gen}^{xy(xz)}$ phases is markedly sharper and more pronounced, suggesting a discontinuous, first-order phase transition.

Fig. \ref{fig:j2d}(b) displays the classical phase diagrams for the $J_1^\Delta K\Gamma\Gamma'J_3$ and $J_1^\Delta K\Gamma\Gamma'J_3^\Delta$ models. The magnetic phases identified are congruent with those delineated for the $J_1K\Gamma\Gamma'J_3^\Delta$ variant in Fig. \ref{fig:j2d}(a). This indicates that the incommensurate spiral order, characterized by an intermediate $\Gamma-{\rm M}$ wavevector, maintains robust stabilization across both the partial and full XXZ renditions of the Hamiltonian when $J_2 = 0$. Broadly, our findings underscore the critical influence of bond-dependent anisotropic exchange interactions ($K, \Gamma, \Gamma'$) in establishing the stability window of these incommensurate spiral phases, thereby ensuring their persistence in parameter space relative to conventional honeycomb $J_1-J_2-J_3$ XXZ models that do not incorporate such directional couplings \cite{RASTELLI:1979}.


%


To systematically elucidate the role of the second-nearest-neighbor coupling $J_2$ in governing the magnetic stability of the modulated states within the full-XXZ $J_1^\Delta K\Gamma\Gamma'J_3^\Delta$ model, we explore the classical phase landscape resolved in the $J_3$--$J_2$ plane using the ab initio baseline of BCAO. As illustrated in Figs.~\ref{fig:j3j2}(a) and \ref{fig:j3j2}(b), the calculations are carried out at two fixed representative spin anisotropy limits: the isotropic Heisenberg limit ($\Delta = 1$) and the pure planar configuration ($\Delta = 0$), respectively. Our LT energy minimization demonstrates that $J_2$ acts as a powerful tuning parameter, driving rich competitions between distinct commensurate regimes and highly tunable incommensurate spin configurations.

The explicit inclusion of a finite $J_2$ term significantly modifies the Fourier-transformed exchange matrix, inducing a pronounced shift in the absolute energy minima and altering the trajectories of the magnetic propagation vector. A salient feature of this updated landscape is the manifestation of the generic incommensurate spiral phase ($I_{\text{gen}}$), which occupies a prominent intermediate region in the $J_3$--$J_2$ plane (as depicted in Fig.~\ref{fig:j3j2}(a)). While the $I_{\Gamma\rightarrow {\rm M}}$ phase remains restricted to its one-dimensional high-symmetry track, the energy minimum within the $I_{\text{gen}}$ regime allows $\mathbf{Q}$ to migrate continuously throughout the two-dimensional interior of the FBZ. Driven by the joint variation of $J_3$ and $J_2$, this unpinned wavevector traces a highly non-trivial trajectory in reciprocal space, reflecting a complex spatial modulation of the underlying planar spin textures. Remarkably, for finite $J_2 > 0$, our calculations also resolve narrow pockets where incommensurate structures become stabilized with their propagation vectors directed along the $\Gamma \rightarrow {\rm K}$ ($I_{\Gamma\rightarrow {\rm K}}$) and ${\rm M} \rightarrow {\rm K}$ ($I_{{\rm M}\rightarrow {\rm K}}$) high-symmetry segments, further demonstrating the rich landscape of modulated instabilities.

A closer inspection of the phase boundaries underscores the decisive role of $J_2$ in shifting the magnetic stability of these modulated configurations. In the $J_2 > 0$ antiferromagnetic regime, the second-nearest-neighbor coupling acts in synergy with the bond-dependent off-diagonal $\Gamma$ and $\Gamma'$ interactions. This cooperative effect widens the stability window of the incommensurate phases, while systematically restricting the commensurate Stripy and ZZ orders to higher $J_3$ boundaries. Conversely, introducing a ferromagnetic second-nearest-neighbor coupling ($J_2 < 0$) triggers a severe energetic competition.
Specifically, across a broad interval of intermediate $J_3$ strengths, the classical ground-state energy per site of the $I_{\Gamma\rightarrow {\rm M}}$ spiral phase exhibits a substantial reduction, achieving a quasi-degeneracy with the uniform FM phase. Our global energy minimization confirms that the energy profiles of these competing magnetic states remain exceptionally close, maintaining a narrow energy split on the order of $\Delta e_g \sim 0.1\,|J_1|$ per site within this frustrated sector.
This energetic proximity throughout the $J_2 < 0$ region provides quantitative evidence for the profound flatness of the classical energy landscape. Consequently, the spatial modulation of the magnetic propagation vector becomes marginally stable and highly susceptible to minute perturbations in the underlying exchange channels, resulting in a delicate classical regime where the modulated spiral and uniform FM phases track each other quasi-identically in energy.

The evolution of these competing configurations is resolved in Figs.~\ref{fig:j3j2}(c) and \ref{fig:j3j2}(d) through a representative one-dimensional cut at fixed $J_2 = 0.4$ and $\Delta = 0.8$. The continuous migration of the energy-minimizing wavevector along the high-symmetry tracks, as well as its departure into the two-dimensional interior of the FBZ, is tracked as a function of $J_3$ in Fig.~\ref{fig:j3j2}(c). Concurrently, Fig.~\ref{fig:j3j2}(d) displays the classical ground-state energy per site, $e_g$ alongside the magnetic susceptibility, $\chi_{J_3} = -\partial^2 e_g / \partial J_3^2$. The precise correspondence between the anomalous peaks in $\chi_{J_3}$ and the boundaries of the modulated regimes confirms the sharp reconfigurations of the underlying spin textures, thereby establishing the exact phase limits driven by the competing exchange channels.

A closer inspection of the phase landscapes in Figs.~\ref{fig:j3j2}(a) and \ref{fig:j3j2}(b) reveals the selective impact of spin anisotropy $\Delta$ on the stability of both modulated and commensurate regimes. Notably, the primary stabilization window of the $I_{\Gamma\rightarrow {\rm M}}$ phase remains largely unaffected, indicating the robustness of wavevector directional locking along the zone-center-to-zone-boundary track against spin-space anisotropy. Conversely, tuning the system to the pure planar limit ($\Delta \rightarrow 0$) entirely destabilizes the commensurate Stripy order, eradicating its presence across the entire $J_3$--$J_2$ plane. Concurrently, this reduction to the pure $XY$ configuration lifts the symmetry constraints within the narrow incommensurate tracks localized at $J_2 > 0$, inducing a local crossover where these line-locked pockets are supplanted by the generic $I_{\text{gen}}$ spiral phase. These distinct features suggest that while off-diagonal exchange ($\Gamma, \Gamma'$) is the fundamental driver for wavevector unpinning, spatial anisotropy $\Delta$ plays a dual role: it is strictly required to protect the commensurate Stripy phase against the generic $I_{\text{gen}}$ spiral phase, while simultaneously enforcing the directional stability of modulated textures within narrow, frustrated sectors.

\section{Quantum Phase Diagram and Incommensurate Spin Structures}
\label{sec:quantum_phase_diagram}

\begin{figure}[t]
	\includegraphics[width=3.30in]{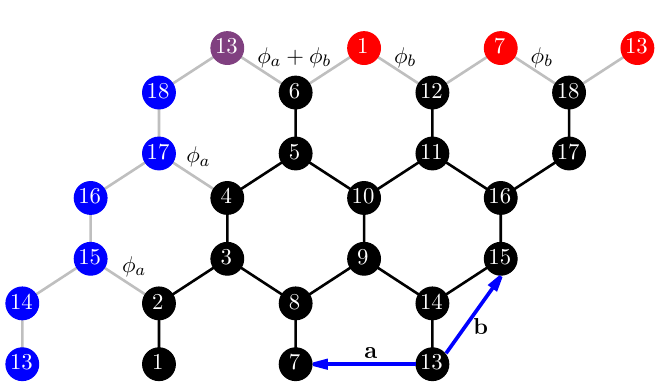}
	\caption{
		(Color online)
This schematic illustrates the $N = 18$ site honeycomb lattice cluster utilized in our finite-system ED calculations. The core physical cluster comprises sites labeled 1 through 18 (black nodes). To simulate an infinite continuum and mitigate finite-size momentum quantization artifacts, TBCs are implemented. Gray links denote boundary couplings, across which phase factors $\phi_a$, $\phi_b$, and $\phi_a + \phi_b$ are accrued during single-particle or spin-multiplet hopping events. Primitive lattice translation vectors are indicated by blue arrows $\mathbf{a}$ and $\mathbf{b}$. Blue, red, and purple nodes represent the boundary image sites, which are mapped via their respective translational fluxes.
	}
	\label{fig:lattice}
\end{figure}

To explore the quantum phase diagram of the extended $J_1\text{--}K\text{--}\Gamma\text{--}\Gamma'\text{--}J_2\text{--}J_3$ Hamiltonian, we perform ED calculations on an $N=18$ site honeycomb cluster [Fig.~\ref{fig:lattice}]. In finite-size geometries, conventional periodic boundary conditions inherently discretize the allowed momentum modes. This often leads to artificial commensurate lock-in or spurious stabilization of magnetic states whose intrinsic propagation vector $\mathbf{Q}$ is incommensurate with the underlying cluster periodicity.


\begin{figure*}[t]
	\includegraphics[width=7.30in]{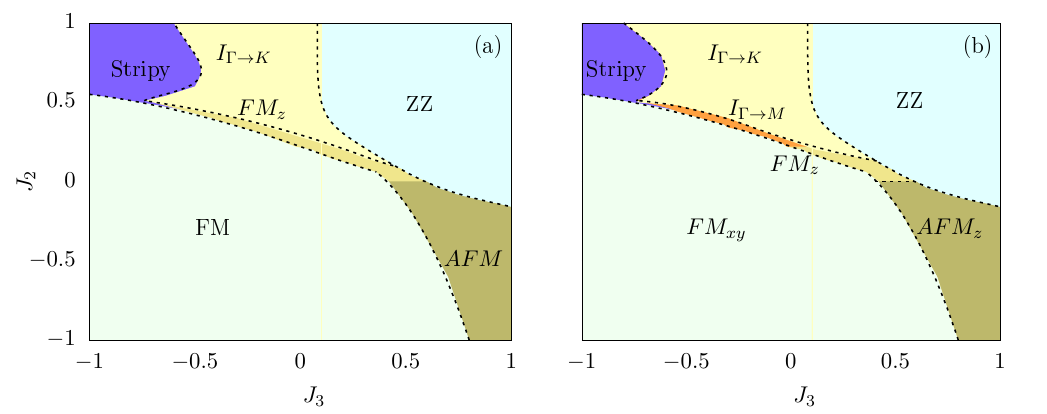}
	\caption{
		(Color online)
		Quantum ground-state phase diagrams of the extended $JK\Gamma\Gamma'$ model on the $N=18$ site honeycomb cluster obtained via TBCs-ED calculations in the $J_3\text{--}J_2$ parameter space. (a) Phase diagram in the isotropic Heisenberg limit ($\Delta = 1$), showing broad domains of collinear phases ($\mathrm{FM}$, $\mathrm{ZZ}$, $\mathrm{AFM}$, Stripy), the axially polarized $\mathrm{FM}_z$ state, and the incommensurate spiral phase $I_{\Gamma\rightarrow K}$. (b) Phase diagram in the easy-plane $XY$ limit ($\Delta = 0$), illustrating the planar $\mathrm{FM}_{xy}$ and axial $\mathrm{AFM}_z$ states, along with the quantum stabilization of the $I_{\Gamma\rightarrow M}$ spiral phase alongside $I_{\Gamma\rightarrow K}$. Dashed black lines indicate quantum phase boundaries determined from energy derivative anomalies.
	}
	\label{fig:qpd}
\end{figure*}

To circumvent finite-size quantization artifacts without perturbing the local exchange Hamiltonian, we implement TBCs by threading magnetic fluxes through the non-contractible loops of the cluster torus along the translation vectors $\mathbf{a}$ and $\mathbf{b}$ [Fig.~\ref{fig:lattice}]. These conditions are parameterized by the boundary phase angles $\boldsymbol{\phi} = (\phi_a, \phi_b) \in [0, 2\pi) \times [0, 2\pi)$ \cite{yuste2017using}. For any exchange link $\langle i,j \rangle$ that crosses the cluster boundary along the translation vector $\mathbf{a}$ or $\mathbf{b}$, the transverse spin operators undergo a gauge transformation $S_j^+ \to S_j^+ e^{i\phi_\mu}$ and $S_j^- \to S_j^- e^{-i\phi_\mu}$ (where $\mu = a, b$), while the $S_j^z$ component remains invariant. The resulting boundary-modulated exchange term is expressed as:

\begin{align}
	\mathcal{H}_{\text{boundary}}(\boldsymbol{\phi}) = \sum_{\langle i, j \rangle_{\text{boundary}}} \sum_{\alpha, \beta} J_{ij}^{\alpha\beta}(\boldsymbol{\phi}) S_i^\alpha S_j^\beta,
	\label{eq:boundary_H}
\end{align}

where $J_{ij}^{\alpha\beta}(\boldsymbol{\phi})$ incorporates the boundary phase factors into the anisotropic exchange tensor, encompassing Heisenberg ($J$), Kitaev ($K$), and off-diagonal symmetric ($\Gamma, \Gamma'$) interactions across the boundary links.

Whereas Ref.~\cite{yuste2017using} employs stochastic sampling over boundary phases to suppress boundary pinning in frustrated spin systems, our approach implements a deterministic and systematic flux-scanning scheme across a fine, uniform mesh within the $[0, 2\pi)^2$ boundary-flux Brillouin zone. This continuous shift of the discrete momentum manifold effectively lifts boundary constraints, enabling the magnetic ordering vector to relax to its true variational minimum. This method serves as an unambiguous diagnostic to distinguish commensurate collinear orders from incommensurate magnetic spirals and featureless quantum disordered regimes. For each flux configuration $\boldsymbol{\phi}$, the exact ground state $|\psi_0(\boldsymbol{\phi})\rangle$ and its corresponding ground-state energy $E_0(\boldsymbol{\phi})$ are obtained to numerical machine precision via the ED method based on the Lanczos algorithm.

\begin{figure*}[t]
	\includegraphics[width=6.60in]{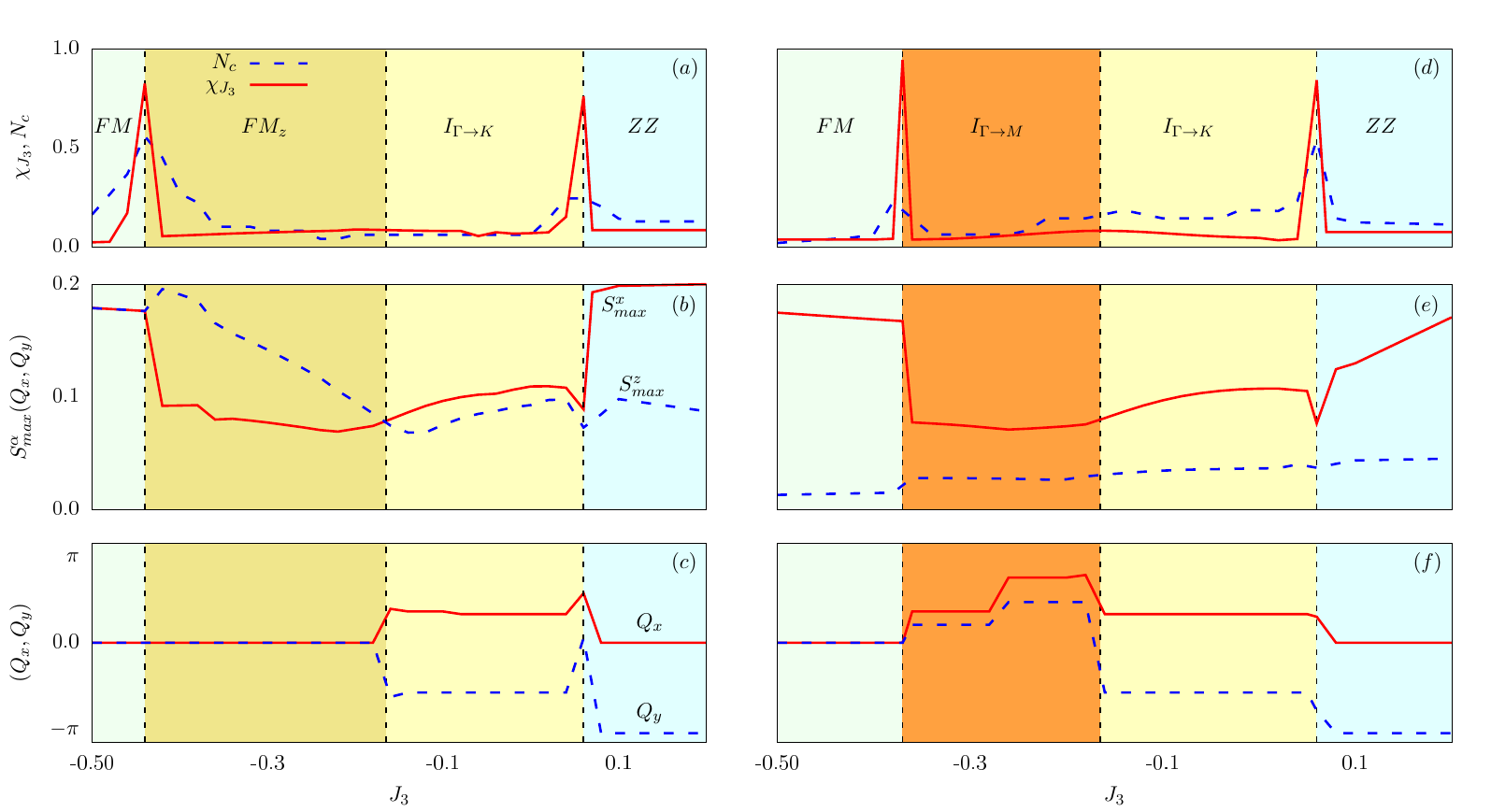}
	\caption{
		(Color online)
	Diagnostic indicators of quantum phase transitions along representative $J_3$ parameter cuts at a fixed $J_2 = 0.4$. 
	The left column displays results for the isotropic Heisenberg limit ($\Delta = 1$), whereas the right column corresponds to the easy-plane limit ($\Delta = 0$). 
	(a),~(d) Effective magnetic susceptibility $\chi_{J_3} = -\partial^2 e_0 / \partial J_3^2$ (solid red curves) and the statistical fluctuation metric $N_c$ (dashed blue curves), where sharp singular peaks unequivocally delineate phase boundaries among competing states ($\mathrm{FM}$, $\mathrm{FM}_z$, $I_{\Gamma\rightarrow M}$, $I_{\Gamma\rightarrow K}$, and $\mathrm{ZZ}$). 
	(b),~(e) Maxima of the static spin structure factor, $S_{\max}^x$ (solid red curves) and $S_{\max}^z$ (dashed blue curves), resolving the in-plane versus out-of-plane spin polarization across crossovers and transition points. 
	(c),~(f) Ordering wavevector components $(Q_x, Q_y)$ associated with the dominant peak of $S(\mathbf{q})$. Continuous evolution of $(Q_x, Q_y)$ tracks the smooth pitch variation of the incommensurate $I_{\Gamma\rightarrow M}$ and $I_{\Gamma\rightarrow K}$ spin spirals, whereas discrete pinning at high-symmetry points marks the commensurate ground states.
}
	\label{fig:sph}
\end{figure*}

\subsection{Characterization of Quantum Disordered and Ordered Magnetic Phases via $N_c$}

\label{subsec:nc_diagnostic}

To differentiate commensurate magnetic ground states from incommensurate spin textures and potential QSL phases, we employ the boundary-flux sensitivity of the cluster, quantified by the diagnostic metric $N_c$. This metric is formulated within the flux-averaging scheme detailed in Ref.~\cite{yuste2017using}.

On a finite cluster, the response of the ED ground-state energy $E_0(\boldsymbol{\phi})$ to a boundary twist $\boldsymbol{\phi} = (\phi_a, \phi_b)$ provides insights into both the spatial spin stiffness and the commensurability of the magnetic order parameter with the imposed boundary conditions. We quantify the density of quasi-degenerate low-lying configurations across the boundary Brillouin zone by defining $N_c$ as:

\begin{equation}
	N_c = \frac{1}{M} \sum_{k=1}^{M} \Theta\left( \varepsilon_{\text{th}} - \frac{E_0(\boldsymbol{\phi}_k) - E_{\text{min}}}{|E_{\text{min}}|} \right),
	\label{eq:Nc_metric}	
\end{equation}

where $E_{\text{min}} = \min_{\boldsymbol{\phi}} E_0(\boldsymbol{\phi})$ represents the absolute ground-state energy obtained over a uniform $M$-point discretization of $\boldsymbol{\phi} \in [0, 2\pi)^2$. Here, $\Theta(x)$ denotes the Heaviside step function, and $\varepsilon_{\text{th}} = 0.01$ establishes an energy threshold of $1\%$.

The boundary-flux diagnostic, $N_c$, offers a quantitative metric for differentiating between disordered and ordered ground states based on their spectral response to flux threading.

{\bf Quantum Disordered and Spin-Liquid Regimes ($N_c \approx 1$)}:
In genuine QSL or resonating-valence-bond states, the absence of dipolar order parameter stiffness and the spatial isotropy of short-range singlets result in a low-lying spectrum that is largely independent of the boundary phase angles. Consequently, the ground-state dispersion, $E_0(\boldsymbol{\phi})$, exhibits global flatness across the entire boundary Brillouin zone, leading to an extensive quasi-degeneracy that drives $N_c \to 1$ \cite{yuste2017using}.

{\bf Magnetically Ordered Regimes ($N_c \ll 1$):}
In states characterized by static magnetic order, the development of a finite order-parameter stiffness robustly pins the ground-state energy, ensuring that $N_c$ remains suppressed well below the spin-liquid threshold. However, the flux-space response demonstrates a distinct topology contingent on momentum commensurability:

\emph{(I) Commensurate Collinear Phases ($N_c \to 0$):} In states locked to discrete high-symmetry momenta commensurate with the cluster (e.g., FM, $\text{FM}_z$, AFM, Stripy, and ZZ phases), boundary twists directly frustrate real-space exchange bonds, imposing a steep spectral gap and isolating the energy minimum, thus yielding $N_c \approx 0$.

\emph{(II) Incommensurate Spiral Phases ($0 < N_c \ll 1$):} For incommensurate magnetic textures (e.g., $I_{\Gamma\to M}$ and $I_{\Gamma\to K}$), the continuous spatial modulation enables the system to partially accommodate flux threading along the modulation direction via an internal shift of the spiral frame. This generates shallow, extended energy valleys in flux space, resulting in a small yet systematically finite baseline ($0 < N_c \ll 1$).

Across the explored parameter space, the diagnostic consistently yields $N_c \ll 1$ and exhibits no indication of an $N_c \approx 1$ plateau, thereby providing conclusive numerical evidence against the realization of a stable zero-field QSL in this model.


\subsection{Quantum Ground-State Phase Diagrams}
\label{sec:quantum_phase_diagrams}

To elucidate the influence of quantum fluctuations on the classical LT ground states of the Kitaev-type model, we conducted ED calculations on an $N = 18$-site honeycomb torus employing TBCs. Figure \ref{fig:qpd} illustrates the resulting quantum phase diagrams within the $(J_3, J_2)$ parameter space. The nearest-neighbor exchange interactions were fixed to values derived from ab initio electronic structure calculations for BCAO ($J_1 = -8.14$, $K = -0.47$, $\Gamma = 2.63$, $\Gamma' = 2.85$~meV). We contrast the isotropic exchange limit ($\Delta = 1$, Fig. \ref{fig:qpd}(a)) with the easy-plane $XXZ$ limit ($\Delta = 0$, Fig. \ref{fig:qpd}(b)).

In the isotropic limit [Fig.~\ref{fig:qpd}(a)], the phase diagram reveals a complex interplay between commensurate and incommensurate magnetic orders. Within the lower-left quadrant ($J_2 \lesssim 0.5$, $J_3 \lesssim 0.4$), the nearest-neighbor exchange $J_1$ predominantly stabilizes the collinear ferromagnetic (FM) state. As frustration intensifies, distinct collinear configurations manifest: an antiferromagnetic $J_2$ coupled with a substantial ferromagnetic $J_3$ favors stripy order ($J_2 \gtrsim 0.5$, $J_3 \lesssim -0.5$), while mutual antiferromagnetic interactions ($J_2 \gtrsim 0.1$, $J_3 \gtrsim 0.15$) drive the system into a robust ZZ phase. Conversely, N{\'e}el antiferromagnetism is restricted to the lower-right sector ($J_2 < 0$, $J_3 \gtrsim 0.4$). In the strongly frustrated intermediate regime ($J_2 \gtrsim 0.5$, $-0.5 \lesssim J_3 \lesssim 0.5$), classical spiral order is quantum mechanically superseded by the incommensurate phase $I_{\Gamma\to K}$, characterized by a pitch vector that evolves continuously along the $\Gamma\to K$ direction. Importantly, quantum fluctuations stabilize an out-of-plane collinear phase, $\text{FM}_z$, within an intermediate corridor adjacent to the FM and incommensurate regimes. This state, devoid of any classical counterpart, arises through a quantum order-by-disorder mechanism, wherein zero-point fluctuations induced by the anisotropic off-diagonal $\Gamma$ and $\Gamma'$ exchanges lift the accidental degeneracy of the classical ferromagnetic manifold, thereby pinning the polarization along the out-of-plane $z$-axis.

In the easy-plane limit [Fig.~\ref{fig:qpd}(b)], the phase diagram undergoes a fundamental restructuring due to the suppression of out-of-plane spin components. The isotropic collinear phases are transformed into the planar ferromagnetic ($\text{FM}_{xy}$) and axial N{\'e}el ($\text{AFM}_z$) states. A notable departure from the isotropic regime is the emergence of the incommensurate spiral phase $I_{\Gamma\to M}$, which propagates along the $\Gamma\to M$ high-symmetry direction and stabilizes below the $I_{\Gamma\to K}$ phase for $J_2 \sim 0.5$. Significantly, this $I_{\Gamma\to M}$ state encroaches upon a portion of the stability region previously occupied by the $\text{FM}_z$ phase. The presence of anisotropic off-diagonal $\Gamma$ and $\Gamma'$ exchanges explicitly breaks the continuous directional degeneracy inherent to the frustrated $J_1\text{--}J_2\text{--}J_3$ spiral manifold, thereby pinning the modulation vector along the $\Gamma\to M$ trajectory. This mechanism offers a robust quantum mechanical explanation for the incommensurate pitch observed experimentally in BCAO.

\subsection{Phase Boundaries and Order Parameters along $J_2 = 0.4$}
\label{sec:diagnostics}

To quantitatively characterize the phase transitions and differentiate between magnetically ordered phases and disordered states, Fig.~\ref{fig:sph} presents an analysis of key diagnostic quantities along a representative parameter sweep at $J_2 = 0.4$ across the range $J_3 \in [-0.5, 0.2]$.

Phase boundaries are delineated by singular peaks in the generalized susceptibility $\chi_{J_3} = -\partial^2 e_0 / \partial J_3^2$, as illustrated in Fig.~\ref{fig:sph}(a) and (d). For $\Delta = 1$, transitions are observed at approximately $J_3 \approx -0.44$, $-0.16$, and $+0.06$, establishing the phase sequence $\text{FM} \rightarrow \text{FM}z \rightarrow I_{\Gamma\rightarrow K} \rightarrow \text{ZZ}$.

To further elucidate the nature of the magnetic ordering, we employ the statistical metric $N_c$, , plotted in Figs.~\ref{fig:fig5}(a) and \ref{fig:fig5}(d), which is defined as the fraction of boundary twist angles $\boldsymbol{\phi}$ for which the ground-state energy $E_0(\boldsymbol{\phi})$ resides within a narrow threshold $\varepsilon_{\text{th}}$ above the global minimum. In commensurate phases---specifically, FM, $\text{FM}_z$, and ZZ---the rigid long-range order is severely penalized by boundary twisting, leading to a vanishingly small $N_c \approx 0$. Conversely, the incommensurate $I_{\Gamma\rightarrow K}$ and $I_{\Gamma\rightarrow M}$ spiral phases exhibit a distinct signature: the gapless sliding phason mode enables the system to accommodate boundary twists through a collective, low-energy shift of the modulation wavevector $\mathbf{Q}$. This manifests as a finite, shallow plateau ($0 < N_c \ll 1$), which serves as a robust diagnostic to differentiate incommensurate spin textures from both rigidly pinned commensurate states ($N_c \approx 0$) and featureless quantum spin liquids ($N_c \sim 1$).

The spin polarization across these transitions is quantified by the peak components of the static structure factor, $S^x_{\max}$ and $S^z_{\max}$ [Figs.~\ref{fig:sph}(b),(e)]. In the isotropic case, $S^x_{\max}$ exhibits an abrupt reduction at $J_3 \approx -0.44$, while $S^z_{\max}$ demonstrates a pronounced enhancement throughout the range $-0.44 \lesssim J_3 \lesssim -0.16$. This observation substantiates the out-of-plane orientation characteristic of the $\text{FM}_z$ phase. Conversely, in the easy-plane limit ($\Delta = 0$), $S^z_{\max}$ is suppressed across the entire parameter space, indicating that magnetic correlations are exclusively confined to the in-plane dimension ($S^x_{\max} > S^z_{\max}$).

Finally, the ordering wavevector $\mathbf{Q} = (Q_x, Q_y)$, determined from the absolute maximum of $S(\mathbf{q})$ [Figs.~\ref{fig:sph}(c),(f)], elucidates the distinction between commensurate lock-in and incommensurate modulation. In the collinear states, $\mathbf{Q}$ is fixed at high-symmetry positions: specifically, $\mathbf{Q} = (0, 0)$ in the FM regime and $\mathbf{Q}={\rm \bf M}$ in the ZZ phase. Conversely, in the intermediate spiral regions ($I_{\Gamma\rightarrow M}$ and $I_{\Gamma\rightarrow K}$), $(Q_x, Q_y)$ exhibits a continuous shift as a function of $J_3$. This continuous evolution reflects the smooth progression of the spiral pitch, which is governed by the interplay between longer-range Heisenberg exchange interactions and anisotropic Kitaev-$\Gamma$ terms.


\section{Discussion and Experimental Implications for $\text{BaCo}_2(\text{AsO}_4)_2$}
\label{sec:discussion}

The magnetic ground state of the layered honeycomb cobaltate BCAO continues to be a subject of considerable debate, primarily concerning whether its low-energy magnetic properties are dictated by bond-directional compass couplings or by spatial exchange frustration~\cite{Halloran:2023,maksimov2025strong,devillez2025bond,Liu:2023}. Early interpretations proposed either strong Kitaev anisotropy~\cite{maksimov2025strong} or the formation of quasi-one-dimensional disordered chains~\cite{devillez2025bond}. However, subsequent elastic and inelastic neutron scattering experiments have revealed an incommensurate magnetic order, characterized by a propagation vector that aligns with high-symmetry directions within the hexagonal Brillouin zone~\cite{Halloran:2023}. More precisely, diffraction profiles exhibit magnetic satellites displaced along the $\Gamma\text{--}M$ path at an incommensurate wavevector of approximately $\mathbf{Q} \approx (0.27, 0)$, thereby establishing a rigorous experimental benchmark for theoretical microscopic spin models.

Our numerical calculations demonstrate that the cooperative interplay of moderate off-diagonal bond anisotropy ($\Gamma, \Gamma'$) and spatial frustration naturally stabilizes this incommensurate spiral manifold along the high-symmetry path, $I_{\Gamma\rightarrow M}$. Crucially, the model captures the experimental ordering pitch without the need to enforce an artificially dominant Kitaev exchange ($|K| \gg |J_1|$). This indicates that moderate bond-anisotropic exchanges, when acting in concert with geometrical frustration, are fully sufficient to account for the zero-field modulation observed in BCAO.

The boundary-twist metric $N_c$ quantifies the spectral rigidity of competing ground states by measuring the fraction of boundary flux sectors $\boldsymbol{\phi}$ for which the ground-state energy $E_0(\boldsymbol{\phi})$ remains within a narrow threshold $\varepsilon_{\mathrm{th}}$ of the global minimum. In commensurate regimes (e.g., $\mathrm{FM}, \mathrm{FM}_z$), where the ordering vector $\mathbf{Q}$ is commensurate with the lattice, boundary twisting incurs a substantial energetic penalty. The absence of low-lying degenerate states results in a stiff energy landscape, effectively pinning the metric to $N_c \approx 0$. Conversely, incommensurate phases ($I_{\Gamma\rightarrow M}$ and $I_{\Gamma\rightarrow K}$) decouple the modulation wavevector from the discrete lattice, thereby conferring a continuous $U(1)$ phase freedom. In these phases, the system accommodates boundary twists via a collective sliding of the magnetic modulation along the cylinder, a process mediated by a gapless phason mode, at a negligible energetic cost. The resultant flatness of the ground-state energy landscape manifests as a robust plateau ($0 < N_c \ll 1$), thereby distinguishing these incommensurate spin textures from both rigidly pinned commensurate states and featureless quantum spin liquids.

\section{Conclusions}
\label{sec:conclusions}

This work elucidates the microscopic origin of the incommensurate magnetic order in BCAO through a combination of LT analysis and ED calculations incorporating TBCs. By anchoring the nearest-neighbor interactions to parameters derived from ab initio electronic structure calculations, we demonstrate that an anomalously dominant Kitaev coupling is not required to reproduce the ground-state phenomenology of BCAO. Instead, the essential physics is governed by the synergistic interplay of moderate bond-directional exchange anisotropies ($\Gamma, \Gamma'$) and spatial frustration arising from farther-neighbor couplings ($J_2, J_3$).

At the classical level, these off-diagonal anisotropies lift the extensive accidental degeneracy of the frustrated isotropic manifold, pinning the coplanar spiral modulation along the high-symmetry $\Gamma\rightarrow M$ direction with a pitch vector matching neutron diffraction observations. Quantum mechanically, the phase diagram harbors a competition between this incommensurate manifold and an out-of-plane collinear $\mathrm{FM}_z$ state stabilized by a quantum order-by-disorder mechanism. A central outcome of our unbiased boundary-flux diagnostics is the unambiguous characterization of the incommensurate floating regimes ($I_{\Gamma\rightarrow M}$ and $I_{\Gamma\rightarrow K}$): the emergence of a finite plateau in the twist metric ($0 < N_c \ll 1$) provides clear spectral evidence for a gapless phason sliding mode that absorbs flux insertions at negligible energetic cost, setting these phases apart from both rigidly pinned commensurate states ($N_c \approx 0$) and genuinely topological spin liquids ($N_c \sim 1$).

In summary, our findings establish a robust microscopic foundation for the zero-field ground state of BCAO, demonstrating that realistic exchange parameters naturally reconcile the observed incommensurate pitch. A crucial next step involves investigating the evolution of this incommensurate spiral manifold under applied in-plane magnetic fields. Such an effort will directly illuminate the microscopic origins of the low critical fields reported experimentally and help clarify the nature of the field-induced intermediate regimes.

\section{acknowledgment}

M.-H.Z. acknowledges financial support from the Office of the Vice Chancellor for Research and Technology at Qom University of Technology (Grant No. G798724). We acknowledge the use of ChatGPT (OpenAI) solely to rephrase text and improve the clarity and flow of the manuscript.

\bibliography{References}

\end{document}